\documentclass[conference]{IEEEtran}
\IEEEoverridecommandlockouts

\usepackage{graphicx}
\usepackage{epsfig,epstopdf,float}
\usepackage{amsmath,amssymb,amsfonts}
\usepackage{delarray}

\newcommand{\ul}{\underline}

\newcommand{\mc}{\mathcal}

\newcommand {\beq} {\begin{equation}}
\newcommand {\eeq} {\end{equation}}
\newcommand {\barr} {\begin{array}}
\newcommand {\earr} {\end{array}}
\newcommand {\bear} {\begin{eqnarray}}
\newcommand {\eear} {\end{eqnarray}}
\newcommand {\bears} {\begin{eqnarray*}}
\newcommand {\eears} {\end{eqnarray*}}

\newtheorem{theorem}{Theorem}
\newtheorem{definition}[theorem]{Definition}

\begin{document}
\title{Mean Field Energy Games in Wireless Networks}

\author{\IEEEauthorblockN{Fran\c cois M\'eriaux\IEEEauthorrefmark{1},
Vineeth Varma\IEEEauthorrefmark{1},\IEEEauthorrefmark{2}
Samson Lasaulce\IEEEauthorrefmark{1}}
\IEEEauthorblockA{\IEEEauthorrefmark{1}Laboratoire des Signaux et Syst\`emes -- LSS (CNRS-SUPELEC-Paris Sud), Gif-sur-Yvette, France \\
Email: \{meriaux,varma,lasaulce\}@lss.supelec.fr}
\IEEEauthorblockA{\IEEEauthorrefmark{2}Orange Labs, Issy Les Moulineaux, France}}

\maketitle

\begin{abstract}
This work tackles the problem of energy-efficient distributed power control in wireless networks with a large number of transmitters. The problem is modeled by a dynamic game. Each transmitter-receiver communication is characterized by a state given by the available energy and/or the individual channel state and whose evolution is governed by certain dynamics. Since equilibrium analysis in such a (stochastic) game is generally difficult and even impossible, the problem is approximated by exploiting the large system assumption. Under an appropriate exchangeability assumption, the corresponding mean field game is well defined and studied in detail for special cases. The main contribution of this work is to show how mean field games can be applied to the problem under investigation and provide illustrative numerical results. Our results indicate that this approach can lead to significant gains in terms of energy-efficiency at the resulting equilibrium.

\end{abstract}

\section{Introduction}
\label{intro}
We study distributed wireless networks, in which mobile terminals have the liberty to choose their own power policies. This can be due to the absence of a central node to control the terminals in the network or due to complexity issues when there is a large number of terminals. Such a scenario can occur with ad hoc networks~\cite{gupta-cdc-1997}, unlicensed band communications, and cognitive radio~\cite{fette-book-2006, mitola-1999}. Modeling the terminals as rational agents, who choose their power control policies to maximize some utilities leads to applying game theory on the problem~\cite{Lasaulce-Tutorial-09,lasaulce-book-2011}. In the problem under study, the wireless network model is a multiple access channel (MAC) and the channel access method is code division multiple access (CDMA).  In this network, the mobile terminals aim to maximize their expected energy-efficiencies over a given time duration. For this purpose, they must adapt their power control policies to varying channel conditions and decreasing energy in their batteries. 

By energy-efficiency, we mean the number of successfully decoded bits at the receiver per Joule consumed at the transmitter, as defined in~\cite{goodman-pc-2000}. In this seminal paper and related works~\cite{meshkati-jsac-2006,bonneau-jsac-2008,lasaulce-twc-2009,buzzi-jstp-2011,Bacci12}, the power control problem is modeled by a sequence of static games independent from one stage to another. But this approach does not capture the interactions that are present among the players when a game is repeated. In~\cite{LeTreustLasaulce(PowerControlRG)10}, it is shown that modeling the problem by a repeated game can lead to more efficient equilibrium power control policies (in the sense of Pareto) than the Nash equilibrium from the static formulation. However, this repeated model uses a normalized stage game which does not depend on the channels realizations. One of the motivations of our work is to account for the impact of channels realizations by modeling the problem with a stochastic game. Moreover, we also account for long-term energy constraint in the terminal since the remaining energy of the battery decreases when power is consumed. Precisely, the energy-efficient power control problem in MAC under long-term energy constraints is modeled by a stochastic differential game (SDG). But the problem of characterizing the performance of distributed networks modeled by SDG becomes hard and even impossible when the number of players becomes large. The same statement holds for determining individually optimal control strategies. In~\cite{meriaux2012}, this problem is overcome by the use of mean field games (MFG). MFG \cite{lasry-jjm-2007} represent a way of approximating a stochastic differential (or difference) game, by a much more tractable model. Under the assumption of individual state information, the idea is precisely to exploit as an opportunity the fact that the number of players is large to simplify the analysis. Typically, instead of depending on the actions and states of all the players, the mean field utility of a player only depends on his own action and state, and depends on the others through an mean field.
The main contribution of this paper is to show that the model developed in~\cite{meriaux2012} can be particularized to cases in which, it is possible to derive equilibrium power control policies. These policies are illustrated with some numerical results and they are compared with the equilibrium from the static game formulation and the equilibrium of the repeated discounted game studied in ~\cite{LeTreustLasaulce(PowerControlRG)10}.

Our paper is structured as follows. We describe the studied wireless network and the evolution laws for the channels and the energy in Sec.~\ref{sec:sys_mod}. In Sec.~\ref{sec: SDG}, the problem of power control is modeled as an SDG and is shown to converge to an MFG under given conditions. Additionally, the regime of large energy budgets and the regime of quasi-static channels are discussed. Resulting equilibrium power control strategies are compared to the classical power control policies from the static game and the discounted repeated game in Sec.~\ref{sec:numeric}.

\emph{Notations:} In the following, $\nabla_{\underline{x}} f$ and $\Delta_{\underline{x}} f$ respectively represent the gradient and the Laplacian of the function $f$ w.r.t. the vector $\underline{x}$. The divergence operator w.r.t. the vector $\underline{x}$ is denoted by $div_{\underline{x}}$. The scalar product in the space $\mathbb{A}$ is represented by $\langle \,,\rangle_{\mathbb{A}}$.


\section{System Model} \label{sec:sys_mod}
We consider one cell with a single base station and $K$ mobile terminals. Since the uplink power control problem is addressed, mobile terminals are transmitters and the base station is the receiver. The radio resource is used as a MAC. Consequently the transmitters interfere with each other. 
Each transmitter sends a signal to a common receiver and has to choose the power level of the transmitted signal. In order to optimize its individual energy-efficiency, i.e., the ratio of its throughput to its transmit power, each transmitter adapts its power level. This choice depends on the quality of the channel between the transmitter and the receiver, the power levels chosen by the other transmitter and the energy remaining. The set of transmitters is denoted by $\mathcal{K} = \{1,\ldots,K\}$ and for each transmitter $k \in \mathcal{K}$, the signal-to-interference plus noise ratio (SINR) writes
 \begin{equation}\label{eq:sinr}
\gamma_k(p_1,\ldots, p_K) = \frac{p_k \left|h_k \right|^2 }{I_k +\sigma^2},
\end{equation}
where $p_k \in \mathcal{P}_k$ and $\left|h_k \right|^2$ are the power level and the channel gain of transmitter $k$, respectively. The variance of the noise is represented by $\sigma^2$. The interference term is denoted by $I_k=\frac{1}{N}\sum_{j \in \mc{K}, j \neq k} p_j |h_j|^2$ with $N$ a processing gain due to interference management at the receiver. For example, in CDMA systems, $N$ represents the spreading factor.

The instantaneous energy-efficiency, in bit/joule is defined as 
\begin{equation}
\label{eq:def-of-utility} u_k(p_1,\ldots, p_K)  = \frac{R
f\left(\gamma_k(p_1,\ldots, p_K) \right)}{p_k} \ [\mathrm{bit} / \mathrm{J}],
\end{equation}
where $R$ is a constant rate of the transmitter. The function $f$ is the probability of having no outage which takes its values in $[0,1]$ and depends on the SINR. 

Two parameters define the state of each transmitter: the channel coefficients to the receiver and the remaining energy in its battery. The evolution of these parameters is modeled by the two following laws. The channel evolution law is
\begin{equation}\label{eq:channellaw}
     \mathrm{d} h_k(t) = \frac12\bigl(\mu - h_k(t)\bigr) \mathrm{d}t + \eta  \mathrm{d} \mathbb{W}_k(t),
\end{equation}
where $\mu \in \mathbb{C}$ and $\eta \in \mathbb{R}$ are constants related to the channel statistics and $\mathbb{W}_k(t)$ are mutually independent Wiener processes.
Depending on the value of $\eta$, this law can model slow-fading or fast-fading. Additional properties about the asymptotic behavior of the channels are given in~\cite{meriaux2012}.

The evolution law of the remaining energy in the battery is given by
\begin{equation}
 \mathrm{d}E_k(t) = - p_k(t) \mathrm{d}t.
\end{equation}
This means that the energy of the battery decreases with the transmit power consumption. 


\section{From the Stochastic Differential Game Formulation to the Mean Field Game}\label{sec: SDG}
\subsection{Stochastic Differential Game}
With the model defined in the previous section, the problem of each transmitter maximizing its expected energy-efficiency over a given time interval can be addressed with an SDG. Time is assumed to be continuous, i.e., $t\in\mathbb{R}$. The time horizon of the game is finite, it is the interval ranging from $T$ to $T'$.
\begin{definition}[SDG model of the power control problem]\label{def-sdg}
The stochastic differential power control game is defined by the $5-$tuple \newline $\mc{G} = \left(\mathcal{K}, \{\mathcal{P}_k\}_{k\in\mathcal{K}}, \{\mc{X}_k\}_{k\in\mathcal{K}}, \{\mc{S}_k\}_{k\in\mathcal{K}}, \{U_k\}_{k\in\mathcal{K}}\right)$ where:
\begin{itemize}
  \item$\mc{K}=\{1,\ldots,K\}$ is the set of players. Here, the players correspond to the transmitters.
\item  $\mc{P}_k$ is the set of actions of player $k \in \mc{K}$. Here, the action set corresponds to the interval of transmit power values. 
  \item $\mc{X}_k$ is the state space of player $k \in \mc{K}$. The game state for player $k\in \mathcal{K}$ at time $t$ is defined by $\ul{X}_k(t) = [E_k(t), h_k(t)]^{\intercal}$.
    \item $\mc{S}_k$ is the set of feedback control policies for player $k \in \mc{K}$. A control policy will be denoted by $p_k(T\rightarrow T')$ which is a function of time between $T$ and $T'$ two reals such that $T'\geq T$;
  \item the average utility function $U_k$ is defined by:
\begin{equation}
\small
U_k\bigl(\ul{p}(T\rightarrow T')\bigr) =   \mathbb{E} \biggl[\int_T^{T'}{u_k(\underline{p}(t), \underline{X}(t)) \text{d}t} + q(\underline{X}(T')) \biggr],
\end{equation}
where $\ul{p}(T\rightarrow T')=\bigl(p_1(T\rightarrow T'),\ldots, p_K(T\rightarrow T')\bigr)$ is the control strategy profile, $\ul{X}(t) = [\ul{X}_1(t),\ldots,\ul{X}_K(t)]$ is the state profile, $q(\underline{X}(T'))$ is the utility at the final state, and $u_k$ is the instantaneous utility.
\end{itemize}
\end{definition}
Even if it can be proven that a Nash equilibrium exists in this game under given conditions~\cite{meriaux2012}, obtaining the expression of an equilibrium requires to solve a system of $2K$ coupled equations. Consequently, the complexity of such a system makes its resolution impossible for $K$ large. This makes us consider the MFG associated with the problem to overcome this complexity issue. 

\subsection{The Mean field game analysis ($K \rightarrow +\infty$)}
\subsubsection{Assumptions}
When the number of players goes to infinity, under the assumption of the exchangeability of the players of the game and the convergence of the interference term, the SDG can be proven to converge to an MFG. The exchangeability property (see~\cite{meriaux2012} for more details) is ensured if each player only knows its individual state and implements an homogeneous admissible control: $p_k(t) = \alpha(t,\underline{X}_k(t))$. A sufficient condition for the convergence of the interference term is $\lim_{K,\,N \to \infty} \frac{K}{N} = \theta > 0$.
In this new game, the set of players is continuous and the generic state of a player is given by $\underline{s}(t) = [E(t), h(t)]^{\intercal}$, whose distribution is given by $m_t$ ($m_t$ is the mean field). The SINR can be rewritten as:
\begin{equation}
\widehat{\gamma}(\underline{s}(t),m_t) = \frac{p(t)|h(t)|^2}{\sigma^2+ \widehat{I}(t,m_t)},
\end{equation}
with $\widehat{I}(t,m_t)$ the interference resulting from the continuum of other players
\begin{equation}
\widehat{I}(t,m_t) = \int_{\underline{s}} |h|^2 \alpha(t,\underline{s}) m_t(\text{d}\underline{s}),
\end{equation}
where $\alpha(t,\underline{s})$ denotes the generic power response at time $t$ and state $\underline{s}$. The instantaneous utility writes
\begin{equation}
\widehat{u}(p(t),\underline{s}(t),m_t)  = \frac{R f(\widehat{\gamma}(\underline{s}(t),m_t))}{p(t)}.
\end{equation}

The main advantage of the MFG formulation is that the utility of each player depends only on its own state $\underline{s}(t)$ and a common mean field $m_t$. 

\subsubsection{Solution to the mean field best-response problem}
For the mean field optimal trajectory $m_t^*$, the best-response of the generic player is such that there exists an average utility:
\begin{equation}\label{eq: mfbr}
\widehat{v}_T = \sup_{p(T \to T')} \mathbb{E} \biggl[  \int_T^{T'} \widehat{u}(p(t),\underline{s}(t),m_t^*) \text{d}t + q(\underline{s}(T')) \biggr].
\end{equation}
Conversely, the control policy resulting from (\ref{eq: mfbr}) has to lead to the distribution trajectory $m_t^*$. Consequently, a solution of the mean field response problem is a solution of the system
\begin{equation}
\label{eq:HJBF-FPK}
\left\{
\begin{aligned}
\frac{\partial \widehat{v}_t}{\partial t} + \tilde{H}(\underline{s}(t),\frac{\partial \widehat{v}_t}{\partial E},m_t)& \\ + \frac12 \langle \mu - h,\nabla_{h} \widehat{v}_t\rangle_{\mathbb{R}^2}   
+\frac{\eta^2}{2} \Delta_{h} \widehat{v}_t&=0, \\
\frac{\partial m_t}{\partial t} + \frac{\partial}{\partial E}(m_t \frac{\partial}{\partial u'} \tilde{H}(\underline{s}(t),\frac{\partial \widehat{v}_t}{\partial E} ,m_t) )& \\
+div_{h}(m_t  \frac12(\mu - h) ) &= \frac{\eta^2}{2} \Delta_{h} m_t,
\end{aligned}
\right.
\end{equation}
with $\widehat{v}_{T'} = q(s(T')),\ $  $m_T$ known and
\begin{equation}
\tilde{H}(\underline{s},u',m) = \sup_p\{ \widehat{u}(p,\underline{s},m) -p.u'\}.
\end{equation}
The first equation is a Hamilton-Jacobi-Bellman-Fleming equation which gives the behavior of $\widehat v_t$ for a given $m_t$. It is coupled with a Fokker-Planck-Kolmogorov equation which gives the behavior of $m_t$ for a given $\widehat v_t$. The former one is a backward equation whereas the latter one is a forward equation. It means that an initial condition for $m_t$ and a final condition for $\widehat v_t$ are needed to solve the system of two equations.

\subsection{Two particular regimes of the mean field game}
Solving (\ref{eq:HJBF-FPK}) in the general case is complicated since the state of a transmitter includes both its energy and its channel coefficients. However, it is possible to study the solutions of the mean field response problem for these two parts separately.
\subsubsection{Large energy budgets}
If energy budgets are large enough, the variation of energy during the game can be neglected. Consequently, only channel coefficients can be considered as the state of the transmitter. In this case, the mean field problem reduces to
\begin{equation}
\left\{
\begin{array}{ll}
\frac{\partial \widehat{v}_t}{\partial t} + \sup_p \widehat{u}( p,h,m_t)=0, \\
\frac{\partial m_t}{\partial t} +div_{h}(m_t  \frac12(\mu - h) ) = \frac{\eta^2}{2} \Delta_{h} m_t.
\end{array}
\right.
\end{equation}
The first equation amounts to choosing the Nash equilibrium from~\cite{goodman-pc-2000} as the power control. The second equation only depends on the channels statistics and give the evolution of the distribution of the channels. The second equation is solved first and the solution $m_t$ is inserted in the first equation to obtain the power control.

\subsubsection{Quasi-static channels}
Considering only energy as the state of a transmitter amounts to assume that the channel coefficients are constant during the time interval $[T,T']$. In this case, the mean field problem turns into
\begin{equation}
\left\{
\begin{array}{ll}
\frac{\partial \widehat{v}_t}{\partial t} + \sup_p\{ \widehat{u}(p,m_t) -p\frac{\partial \widehat{v}_t}{\partial E}\} =0, \\
\frac{\partial m_t}{\partial t} - \frac{\partial}{\partial E}(m_t p^*) = 0, \\
\end{array}
\right.
\end{equation}
with $p^* = \arg \sup_p\{ \widehat{u}(p,m_t) -p\frac{\partial \widehat{v}_t}{\partial E}\}$. 
The first equation gives $\widehat v_t$ and $p^*(T \to T')$ given $m_t$. The second equation gives $m_t$ given $p^*(T \to T')$. Equilibrium power control policy resulting from this case is illustrated in the following section. 


\section{Numerical Results for a quasi-static channels}\label{sec:numeric}

In this section, we provide illustrative numerical results for the particular case of quasi-static channels, i.e., when the energy dynamics are faster than the channel dynamics. For the implementation of the proposed scheme, each terminal requires only the knowledge of its own channel state and energy level to choose the transmit power.

\subsection{Comparison with other NE}
For the purpose of evaluating our results, we compare the equilibrium of the MFG to the equilibrium of two other well known games.
\begin{enumerate}
\item The static Nash: This is the classical NE from the work of~\cite{goodman-pc-2000}. The equilibrium point is given by the equation:
 \begin{equation}
 \forall k \in \mathcal{K}, \
p_k^{*}= \frac{\sigma^2}{\left|h_k \right|^2} \frac{\beta^*}{1-\theta\beta^{*}}
\label{eq:NE-power}
\end{equation}
where $\beta^*$ is the unique solution of the equation
\begin{equation}\label{eq: nash oneshot}
 xf'(x)-f(x)=0.
\end{equation}
\item The repeated game "operating point": When the power control game is treated as a discounted repeated game, there are several NE. In \cite{LeTreustLasaulce(PowerControlRG)10}, the authors propose an equilibrium point, defined as the ``operating point" which can be very close to the global optimal point. This equilibrium is given by the equation:
 \begin{equation}
 \forall k \in \mathcal{K}, \
\tilde p_k= \frac{\sigma^2}{\left|h_k \right|^2} \frac{\tilde\beta}{1-\theta\tilde\beta}
\label{eq:NE-power}
\end{equation}
where $\tilde\beta$ is the unique solution of the equation
\begin{equation}\label{eq: repeated game}
 x\bigl(1-\theta x\bigr)f'(x)-f(x)=0.
\end{equation}
\end{enumerate}

\subsection{Parameters used}
For the purpose of simulations, in order to obtain useful numerical results we take the following parameters:
\begin{enumerate}
\item the rate $R=1$ Mbps,
\item the noise level with path loss, $\sigma^2=0.1$ W,
\item the channel gain $\mathbb{E} |h(t)|^2 =1$,
\item the success function $ f(\gamma) = \exp ( -\frac{0.9}{\gamma})$~\cite{belmega-tsp-2011}.
\end{enumerate}
The averaged channel gain for all players at all times is taken to be a constant $\mathbb{E} |h(t)|^2 =1$.
The initial distribution in energy of the terminals  $m_T$ is specified for each figure. The final condition $\hat{v}_{T'} = q(s(T'))$ is set to $0$, in order for the comparison to be fair with equilibrium power policies from the static game and the repeated discounted game. Indeed, there is no final reward in these two games. We take the maximum available energy to be $E_{\max}=20$ J and time duration $T'-T$ is specified for each figure.

\begin{table}
\begin{tabular}{|r|l|l|r|}
  \hline
Fig. No& $T'-T$ & $m_t(E,t=0)$ & Plotting\\
  \hline
  \ref{fig:pph1} & 20 & 1 & $P^*$ v.s $(t,E)$ \\
  \hline
 \ref{fig:pph2} & 20 & 1 & $P^*$ v.s $t$ \\
  \hline
 \ref{fig:em1} & 20 & 1 & $m_t$ v.s $(t,E)$\\
  \hline
 \ref{fig:em2} & 20 &  $0$ if $E \leq 18$, else $1$ & $m_t$ v.s $(t,E)$\\
  \hline
 \ref{fig:pph3} & 120 & 1 &  $P^*$ v.s $t$\\
  \hline
 \ref{fig:comp} & 120 & 1 &  $v_t(E,t=0)$ v.s $E$ \\
  \hline
\end{tabular}
\caption{Simulation parameters for the presented figures.}
\end{table}

\subsection{Discussion}

\begin{figure}[H]
    \begin{center}
        \includegraphics[width=90mm]{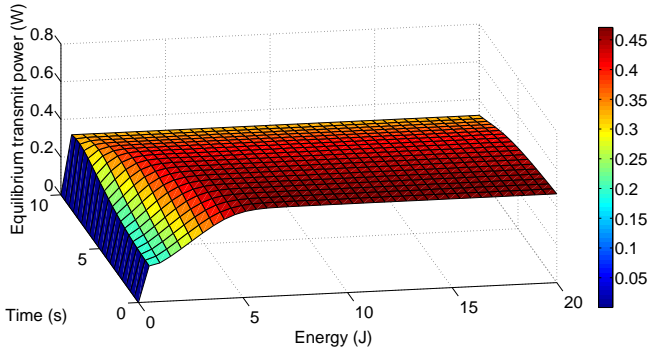}
 
 \end{center}
   
\caption{MFG equilibrium power policy w.r.t. time and energy. \label{fig:pph1}  }
\end{figure}

\begin{figure}[H]
     \begin{center}
         \includegraphics[width=100mm]{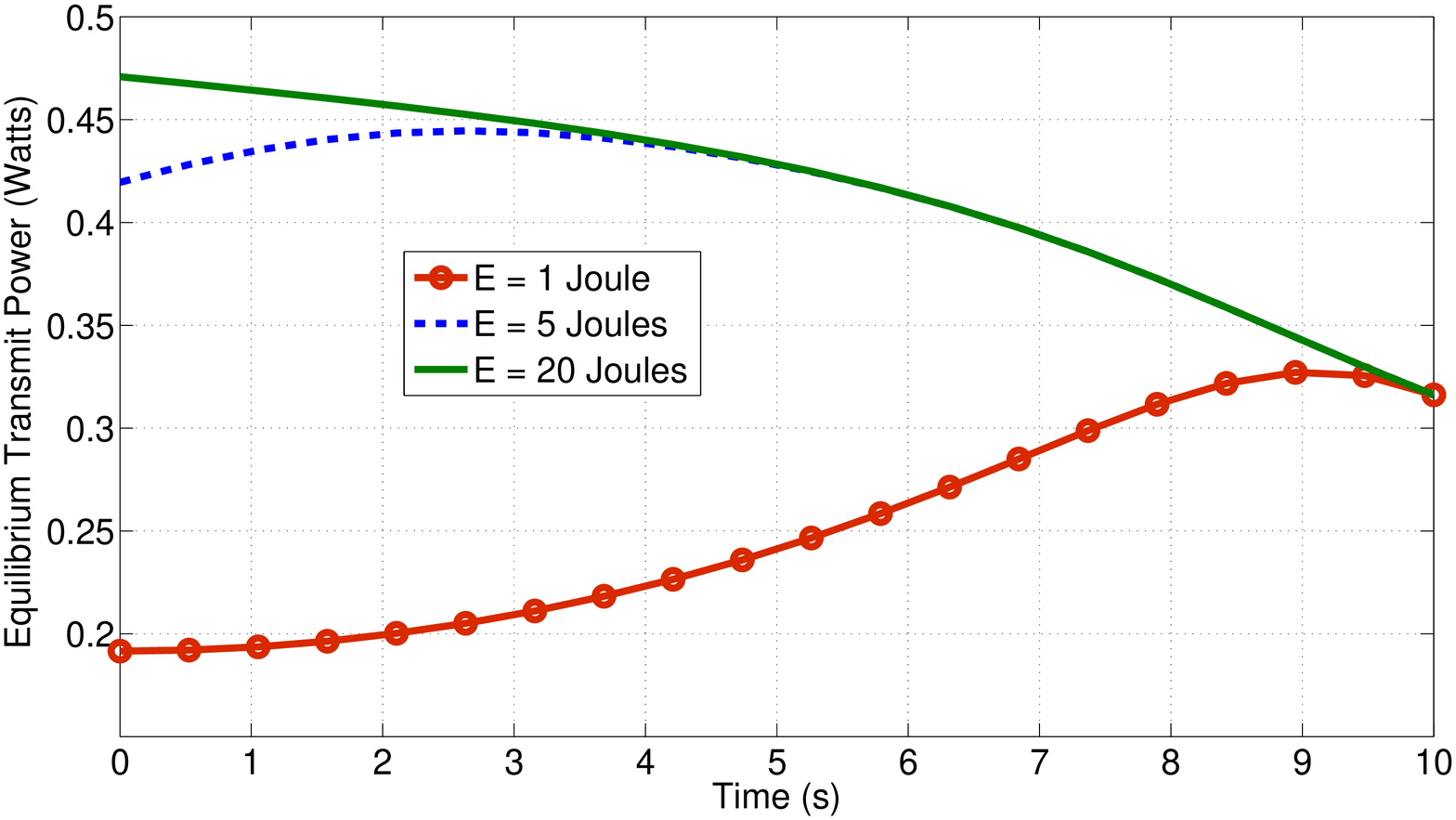}
     \end{center}
   
 \caption{MFG equilibrium power policies w.r.t. time for $3$ initial amounts of energy, for a short game duration.  \label{fig:pph2}}
 \end{figure}

On Fig.~\ref{fig:pph1}, the equilibrium power policy plot shows that terminals starting the game with high energy level ($20$ J in the figure) start transmitting with a high power level. Then this power decreases with time.  It can also be noted that terminals starting the game with low energy do not transmit at high power values at the beginning of the game. They first wait for other terminals to empty their battery in order to suffer from less interference. This phenomenon is highlighted on Fig.~\ref{fig:pph2}. 

\begin{figure}[H]
    \begin{center}
        \includegraphics[width=90mm]{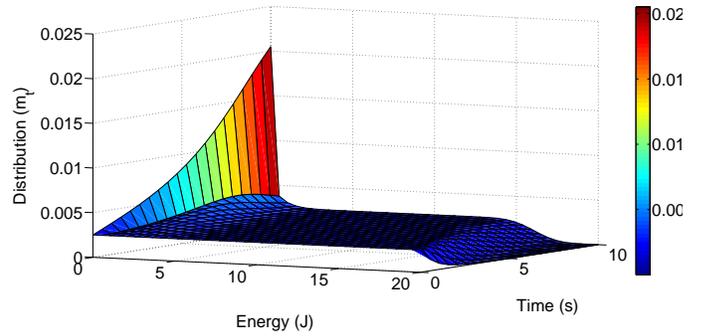}
    \end{center}
   
\caption{Evolution of the energy distribution with the equilibrium power policy of the MFG for uniform initial distribution. \label{fig:em1}}
\end{figure}

\begin{figure}[H]
     \begin{center}
         \includegraphics[width=90mm]{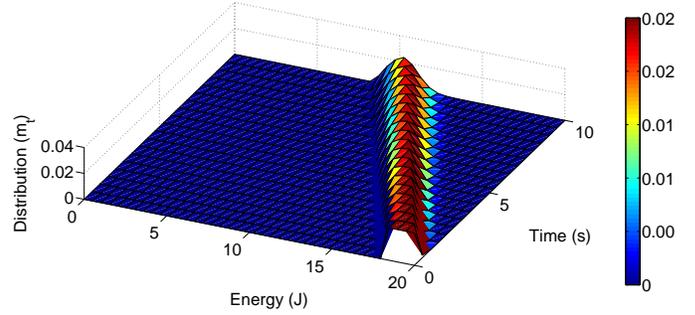}
     \end{center}
  
 \caption{Evolution of the energy distribution with the equilibrium power policy of the MFG for non-uniform initial distribution.   \label{fig:em2}}
 \end{figure}

On Fig.~\ref{fig:em1}, where terminals start the game with a uniform energy distribution, it can be seen that the amount of terminals with high energy ($20$ J) decreases with time, whereas an increasing proportion of terminals have empty batteries. A similar behavior is captured on Fig.~\ref{fig:em2}, except that there are only terminals with high energy at the beginning of this case. 

 \begin{figure}[H]
     \begin{center}
         \includegraphics[width=100mm]{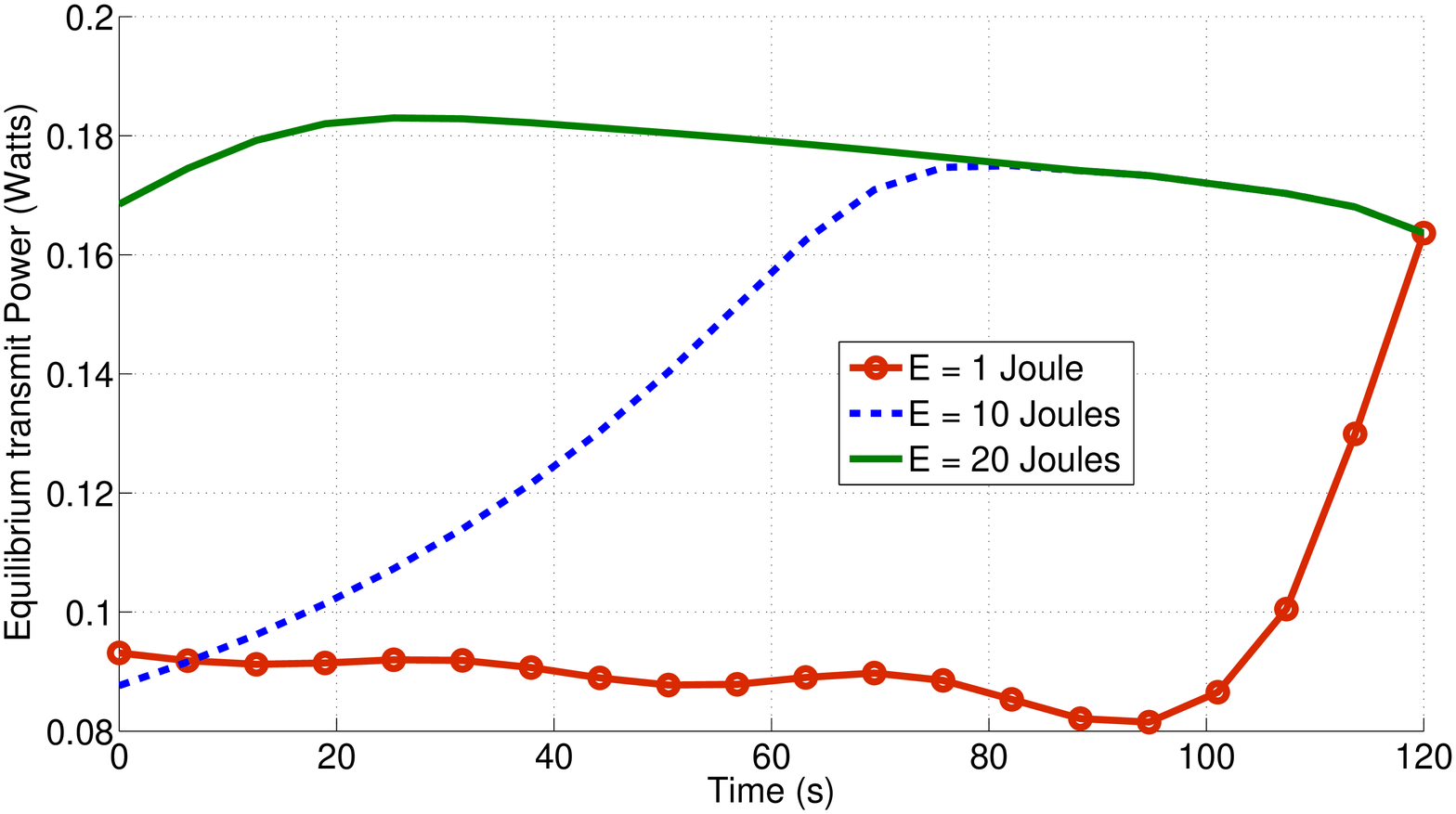}
     \end{center}
    
 \caption{MFG equilibrium power policies w.r.t. time for $3$ initial amounts of energy, for a long game duration.  \label{fig:pph3}}
 \end{figure}

\begin{figure}[H]
    \begin{center}
        \includegraphics[width=100mm]{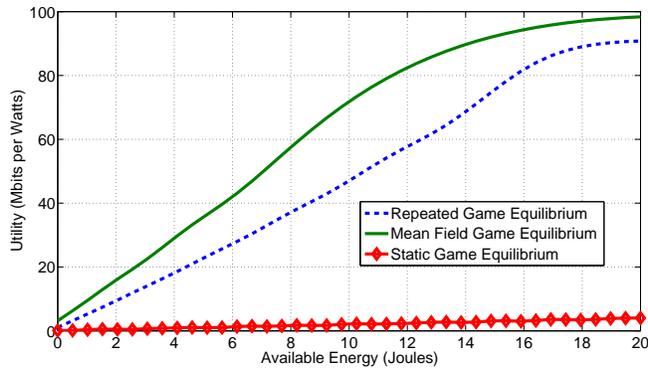}
    \end{center}
    
\caption{Comparison of the average energy-efficient utilities with the MFG equilibrium power policy, for a long game duration, against the equilibrium from the static game, and the operating point from the repeated game formulation defined earlier. \label{fig:comp}}
\end{figure}

Since the solution to the mean field response problem is determined for a given time duration, the equilibrium power control policy naturally depends on this duration. Hence, it is interesting to increase the time duration, as it is done on Fig.~\ref{fig:pph3}, to capture the impact on the power policy. Globally, it entices the terminals to consume less power to make their batteries last longer. This low consumption gives the terminals a better energy-efficiency than with the equilibrium form the static game or the equilibrium from the discounted repeated game, as illustrated on Fig.~\ref{fig:comp}.

\section{Conclusion}

In this work, we treat the problem of power control games as a mean field game, taking into account the limited energy available to mobile transmitters, and the effects of channel fading. Provided that convergence conditions are met, the mean field approach offers interesting and useful results. While in the general case, the resulting partial differential equations are hard to solve, we analyze some special cases where numerical results can be obtained. The numerical results are encouraging, but they have only been computed for the case of quasi-static channels.

Some of our key results that can be interesting to engineers and other researchers are presented below:
\begin{enumerate}
\item Terminals with a low starting energy transmit with lower power at the start and slowly increase their power with time, when the energy distribution is homogeneous and $T'-T$ small.
\item In the same case, terminals with a high starting energy start with a high power and slowly decrease their power with time.
\item Terminals with a low starting energy transmit with lower power at the start and slowly decrease their power with time and then raise, when the energy distribution is homogeneous and $T'-T$ large.
\item The equilibrium of the mean field game when $T'-T$ is large, outperforms other known equilibrium (from \cite{goodman-pc-2000} and \cite{LeTreustLasaulce(PowerControlRG)10}) in terms of energy-efficiency.
\end{enumerate}

A numerical solution to the general case of an evolving channel as well as the energy constraint, is a possibility for future work. Additional possibilities include considering a multi-cellular network, distributed base stations and the multi-carrier case.


\bibliographystyle{abbrv}
\bibliography{biblio-book-2011-03-13-bis}

\end{document}